\documentstyle[floats,twocolumn,aps,psfig,prl]{revtex}

\begin{document}
\draft

\twocolumn[\hsize\textwidth\columnwidth\hsize\csname @twocolumnfalse\endcsname

\title{Anomalous temperature dependence of magnetic quantum oscillations
in CeBiPt}

\author{G. Goll$^1$\cite{grenoble}, J. Hagel$^1$\cite{grenoble},
H. v. L\"ohneysen$^1$, T. Pietrus$^1$, S. Wanka$^1$, J. Wosnitza$^1$\\
T. Yoshino$^2$, T. Takabatake$^2$}

\address{$^1$Physikalisches Institut, Universit\"at Karlsruhe,
D-76128 Karlsruhe, Germany\\
$^2$Department of Quantum Matter, ADSM, Hiroshima University,
Higashi-Hiroshima 739-8526, Japan
}

\date{\today}
\maketitle

\begin{abstract}
Shubnikov--de Haas (SdH) and Hall-effect measurements of CeBiPt and
LaBiPt reveal the presence of simple and very small Fermi surfaces
with hole-like charge carriers for both semimetals. In the magnetic
material, CeBiPt, we observe a strong temperature dependence of the
SdH frequency. This highly unusual effect might be connected with
an internal exchange field within the material and a strongly
spin-dependent scattering of the charge carriers.

\end{abstract}
\pacs{PACS numbers: 71.20.Eh, 71.18.+y}
\vskip2pc]

The Shubnikov--de Haas (SdH) effect, i.e., quantum oscillations in the 
magnetoresistivity $\rho(B)$, was discovered about 70 years ago in 
the semimetal Bi \cite{sdh1930}. Since then, innumerable studies on
the SdH effect and the related de Haas--van Alphen (dHvA) effect in the 
magnetization were reported for many metals, metallic compounds, and 
semimetals. Both effects arise from the oscillations of the free
energy of the electrons as with increasing field $B$ a sudden change of
population of Landau cylinders occurs whenever a Landau cylinder is
pushed through an extremal cross-section $A$ of the Fermi surface
perpendicular to the magnetic field. The oscillations are periodic
in 1/$B$ and the frequency $F$ is directly proportional to $A$, i.e.,
$F = (\hbar/2\pi e)A$. The theory of the dHvA effect is well established
\cite{dhvatheory}, and also the SdH effect, reflecting the more involved
transport properties, is principally understood \cite{sdhtheory}.
In recent years, SdH and dHvA effects have been exploited to establish the
Fermi surface of complex materials with strong electron correlations such
as heavy-fermion systems \cite{dhvasces}. In many cases, the effective mass
of the carriers in a given band of the Fermi surface can be determined from 
the temperature dependence of the amplitude of the oscillations. Here we
report on a new observation, namely a strongly temperature-dependent
{\it frequency} of the SdH oscillations. This anomalous $T$ dependence
of $F$ was found for a certain field orientation in semimetallic CeBiPt,
while it is absent in the homologous - except for the Ce 4$f$ electron
- semimetal LaBiPt.

The single crystals were grown at Hiroshima University with the Bridgman
technique in a hermetically sealed Mo crucible from the starting materials
Ce (m5N, Ames Laboratory), Bi (m5N), Pt (m3N), and La (m4N). To avoid
oxidation of elemental
Ce or La during handling, CePt or LaPt were first prepared by argon-arc
melting and the appropriate amount of Bi then was added for single-crystal
growth. The crucible, sealed under Ar atmosphere, was heated to
1350$^{\circ}$C in a furnace (with an intermediate
halt at 500$^{\circ}$C for 2\,h) and after 12 hours was slowly cooled
by moving it out of the central zone of the furnace with 1\,mm/h. This
cooling process from 1350$^{\circ}$C to 20$^{\circ}$C took 6 days, then
the furnace was switched off. Whether the 1-1-1 compounds or 3-4-3
compounds like Ce$_3$Bi$_4$Pt$_3$ are formed, depends very sensitively
on the excess amount of Bi. The 1-1-1 compounds show the F$\bar 4$3m cubic
structure previously determined for polycrystalline samples \cite{can91}
with no indications of second phases. The lattice constant
6.8338\,(25)$\,{\rm \AA}$ found for CeBiPt is in good agreement with 
the previously reported value \cite{can91}. For the LaBiPt crystal,
Laue-diffraction pictures showed some mosaicity of the sample.

The longitudinal and transverse magnetoresistance (Hall effect) were 
measured with a standard $^3$He cryostat setup up to 15\,T in Karlsruhe, 
and up to 28\,T at the High Magnetic Field Laboratory in Grenoble. Both
sets of data agree with each other in region of overlap. Six gold wires
were glued with graphite paste to the samples, thereby enabling to measure
simultaneously the longitudinal ($\rho_{xx}$) and transverse ($\rho_{xy}$)
magnetoresistance. These resistances were measured by use of a
low-frequency ($\sim 16$\,Hz) lock-in technique for normal and reversed
field orientations which allowed a well-defined separation of $\rho_{xx}$
and $\rho_{xy}$. The dHvA signal of the LaBiPt sample was measured by
means of a capacitance cantilever torque magnetometer. All sample
holders could be rotated {\it in situ} around one axis.

\begin{figure}[ht]
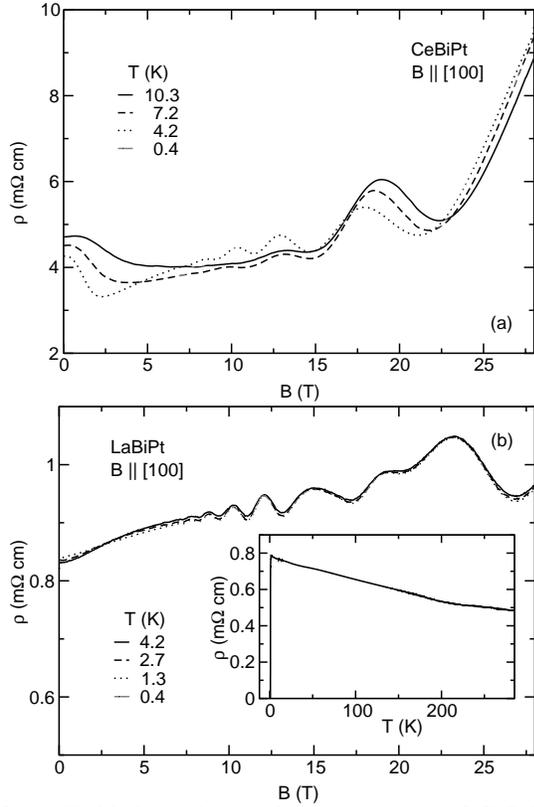

  \centerline{\psfig{file=goll_f1a.eps,clip=,width=7cm}}
  \centerline{\psfig{file=goll_f1b.eps,clip=,width=7cm}}
\caption[mr]{Field dependence of the resistivity of CeBiPt (a)
and LaBiPt (b) for selected temperatures. The inset shows the
temperature dependence of the resistance of LaBiPt with the
superconducting transition at about 0.88\,K.}
\label{mr}
\end{figure}

Fig.\ \ref{mr}(a) and (b) show the magnetoresistance $\rho(B)$ for
CeBiPt and for LaBiPt with field along [1\,0\,0] for several
temperatures. The current direction was perpendicular to $B$. In both
cases, clear oscillations are seen. For LaBiPt at $T$ = 0.4\,K, the
strong decrease of $\rho$ below about 1\,T (with a 50\% value at
0.45\,T) indicates the onset of superconductivity with $T_c$ =
0.88\,K [see inset of Fig.\ \ref{mr}(b) for $\rho(T)$]. The rather large
critical-field slope (${\rm d}B_{c2}/ {\rm d}T)_{T_c} \approx $
-1\,T/K hints at a small coherence length. For some field
directions we observed a beat pattern for LaBiPt which might be attributed
to the mosaicity mentioned above. We can, however, not exclude the
existence of a second extremal Fermi surface of small area in LaBiPt.

For CeBiPt, a first unusual observation is the sharp drop of the
magnetoresistance at low fields and low temperatures [Fig.\ \ref{mr}(a)].
Below about $T_N =1$\,K, CeBiPt is in an antiferromagnetically ordered
state as was evidenced by specific-heat and magnetization measurements
for our samples \cite{pie00,remtn}. Below $T_N$, the field-dependent
magnetization, $M$, has a maximum in ${\rm d}M/{\rm d}B$ at about
0.3\,T \cite{pie00}. Therefore, the negative magnetoresistance at
low fields presumably is caused by antiferromagnetic fluctuations
which become reduced in an applied magnetic field.

\begin{figure}[ht]
  \centerline{\psfig{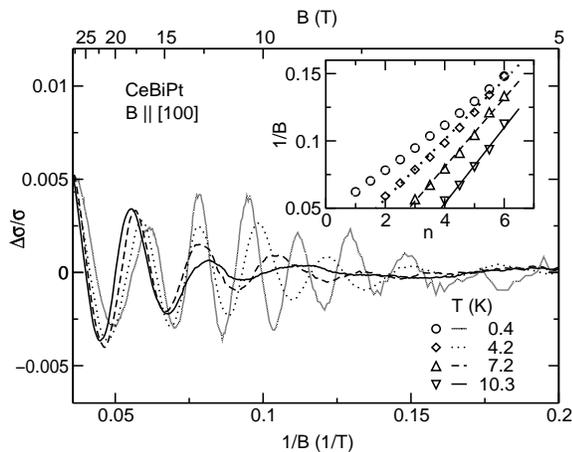}}
\caption[sdh]{Field dependence of the SdH signal, i.e., the relative
conductance oscillations, of CeBiPt for different temperatures. The
inset shows the peak and valley positions vs an arbitrary index $n$.}
\label{sdh}
\end{figure}

The most striking observation for CeBiPt is the shift of the
oscillating SdH signal with temperature. This can be made
more apparent when the resistivity is converted to conductivity via
$\sigma_{xx} = \rho_{xx}/(\rho_{xx}^2 + \rho_{xy}^2)$ making use of the
simultaneously measured transverse magnetoresistance $\rho_{xy}$
\cite{remhall}. A smooth background longitudinal conductivity
$\sigma_0$ can be fitted to the $\sigma_{xx}$ data to obtain the
SdH signal $\Delta\sigma/\sigma_0 = (\sigma_{xx} - \sigma_0)/\sigma_0$
as shown in Fig.\ \ref{sdh}. The inset of Fig.\ \ref{sdh} displays
the sequence of the position of 1/$B$ of the oscillation maxima and
minima.

\begin{figure}[ht]
  \centerline{\psfig{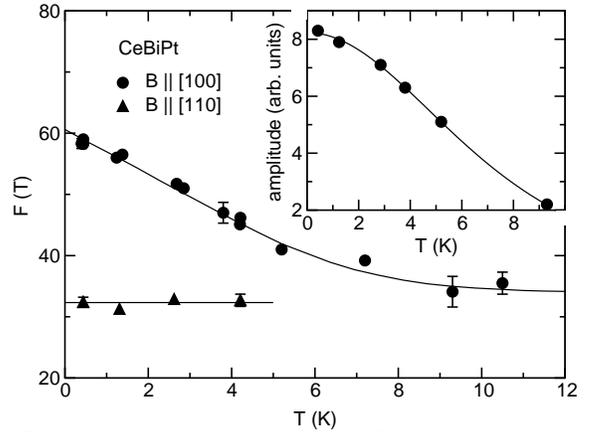}}
\caption[fvst]{Temperature dependence of the SdH frequency of
CeBiPt for two different field orientations. The lines are guides
to the eye. The inset shows
the temperature dependence of the SdH oscillation amplitude with
the theoretical fit for an effective cyclotron mass of $m^\ast =
0.24$\,$m_e$.}
\label{fvst}
\end{figure}

The oscillations are indeed periodic in $1/B$ within our resolution
and for the limited field range. The oscillation frequency $F$ is
directly obtained from the slope of $1/B$ vs.\ oscillation index $n$.
$F$ decreases from 58.2\,T at 0.4\,K to about 35\,T at 10.3\,K.
Additional data taken in Karlsruhe in fields up to 15\,T are fully
in line with this decrease (see Fig.\ \ref{fvst}). The total change
of $F$ between 0.4 and 10.3\,K corresponds to an apparent reduction of
the Fermi-surface cross-section by more than 50\%! The inset of
Fig.\ \ref{fvst} shows the decrease of the amplitude $A$ of the SdH
oscillations with increasing $T$ as measured at 10.5\,T. $A$ can be
described quite well with the standard expression $A \sim T/
\sinh[(14.69 (T/B) (m^\ast/m_0))]$, where $T$ and $B$ are expressed
in K and T, respectively \cite{sho84}.
We obtain the effective mass $m^\ast = 0.24\,m_0$, where $m_0$ is the
free-electron mass. A simple consistency check serves to test whether
we are indeed dealing with a quantum-oscillation phenomenon.
From the field dependence of $\Delta \sigma/\sigma_0$ at fixed $T$ one
can estimate the charge-carrier scattering rate $\tau^{-1}$.
The Dingle temperature $T_D \approx 4$\,K obtained from the
field dependence of the SdH amplitude (at lower fields) corresponds
to $\tau^{-1} = 3.3 \times 10^{12}$\,s$^{-1}$. Together with the carrier
density $n_h = 7.7 \times 10^{17}$\,cm$^{-3}$ obtained from the
Hall effect at liquid-He temperature we obtain in the simple Drude
model $\rho \approx 3.7$\,m$\Omega$cm which is perfectly in line with
the measured resistivity.

\begin{figure}[ht]
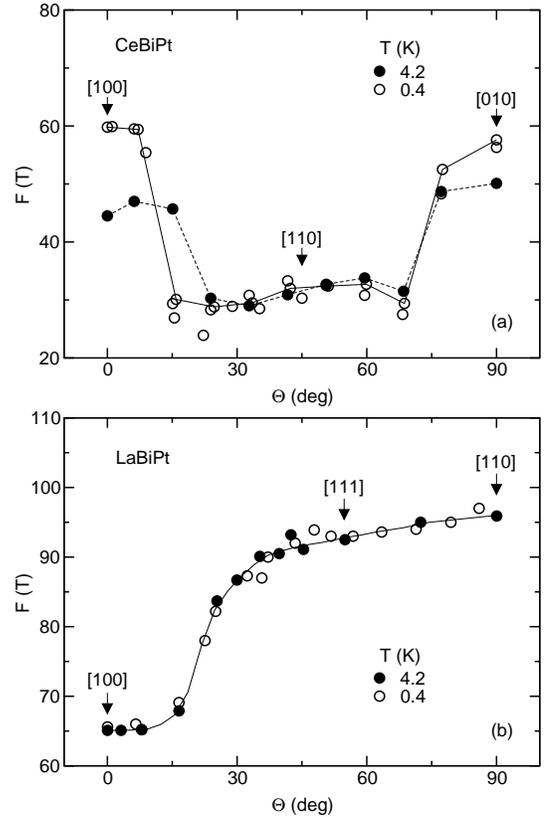

  \centerline{\psfig{file=goll_f4a.eps,clip=,width=7cm}}
  \centerline{\psfig{file=goll_f4b.eps,clip=,width=7cm}}
\caption[fvsang]{Angular dependences of the SdH frequencies of (a)
CeBiPt and (b) LaBiPt for two different temperatures.}
\label{fvsang}
\end{figure}

The angular dependence of $F$ for CeBiPt is depicted in
Fig.\ \ref{fvsang}(a) for $T$ = 0.43\,K and 4.21\,K. The anomalous
$T$ dependence is found only for field along the [1\,0\,0] and
[0\,1\,0] directions where $F$ is found to be somewhat larger,
i.e., about 58\,T. Between approximately $\theta = 20^\circ$ and
$70^\circ$, where $\theta$ is the angle from the [1\,0\,0]
direction, $F$ is constant at 30\,T and furthermore independent of
$T$. For $B$ along the [1\,1\,1] direction a very low SdH frequency
of about 20\,T - but again independent of $T$ - was found.
LaBiPt shows a temperature-independent angular dependence of $F$
[Fig.\ \ref{fvsang}(b)]. The SdH and the dHvA frequency are within
error bars identical and somewhat larger ($F$ increases from 65\,T for
$B$ along [1\,0\,0] to 95\,T for $B$ along [1\,1\,0]) than $F$ for CeBiPt.
The large magnetization of CeBiPt due to the Ce $4f$ electrons
prevented us from taking dHvA data of the torque magnetization for this
compound. Nevertheless, the very good agreement between SdH and
dHvA data for LaBiPt lends support to the SdH data for CeBiPt.
For both materials we were able to observe a SdH signal over the whole
angular range. This suggests that the Fermi surfaces are simple
single-connected hole pockets. The volume enclosed by the Fermi
surface is estimated to comprise only about $1.5\times 10^{-4}$ of
the volume of the first Brillouin zone for CeBiPt. This is consistent
with the low (hole-like) charge-carrier concentration of $n_h =
7.7\times 10^{17}$\,cm$^{-3}$. Within a free-electron picture,
$n_h$ corresponds to a Fermi energy of $E_F =
\hbar^2(3\pi^2 n_h)^{2/3}/2m^\ast = 12.8$\,meV. Assuming a circular
Fermi-surface cross section, i.e., $A = \pi k_F^2$
with the Fermi wavevector $k_F$, this results in a SdH
frequency of $F = m^\ast F/\hbar e \approx 27$\,T in nice agreement
with the experimental $F$ between $20^\circ$ and $70^\circ$.

The main point of the present investigation is the observation
of a temperature dependence of the quantum-oscillation
{\it frequency} which is found only for the Ce-based metal.
To our knowledge such an effect has never been observed before.
As a possible origin for this behavior one could assume a
temperature-dependent charge-carrier density which would lead to
a change in the Fermi-surface topology. However, the simultaneously
measured Hall constant $R_H = 1/n_h e$, is independent of temperature.

Another possibility would be a reconstruction of the Brillouin
zone. This might be caused, e.g., by an antiferromagnetic ordering,
which introduces an extra periodicity into the lattice.
The effect of magnetic ordering has previously been observed in
a {\it field-dependent} change of the dHvA frequency in
NdIn$_3$ \cite{ume92} or in an unusual temperature
dependence of the dHvA {\it amplitude} in YbAs \cite{tak93},
SmSb \cite{tan93}, and SmAgSb$_2$ \cite{mye99}.
As mentioned, CeBiPt undergoes an antiferromagnetic transition at
about 1\,K. However, no abrupt change of $F$ with $T$ around
$T_N$ or $B$ but a rather smooth variation over a large $T$ range
is observed. This indicates that a Brillouin-zone reconstruction
must be ruled out as a cause for the observed frequency change.

One important fact which is evident from our investigation is the
absence of any unusual effect for the non-magnetic sister compound
LaBiPt. Therefore, it is clear that the magnetism of the
Ce atoms affects the magnetic quantum oscillations. For a sample
possessing an internal magnetization, the magnetic
induction $B_i = \mu_0 (H + M)$ is different from the externally
applied field $B = \mu_0 H$. The SdH signal is proportional
to $\sin(2\pi F/B_i)$. Therefore, a temperature-dependent
magnetization would cause a change in the SdH frequency. However,
in order to explain the experimentally observed increase of
the SdH frequency with decreasing temperature, $M$ would
have to become smaller at lower temperatures. This is contrary
to the usual behavior of $M$ and not in line with the measured
magnetization which is about $\mu_0M = 0.2$\,T at $\mu_0H = 12$\,T and
$T = 1.7$\,K and decreases with increasing temperature. Moreover,
the magnitude of $\mu_0M$ compared to $B$ is much too small
to account for the observed change of $F$.

An anomalous situation may occur when we suppose that a
temperature-dependent (and possibly also a field-dependent)
phase difference $\Phi$ exists between the spin-up and
spin-down oscillations and that, in addition, their amplitudes
may be different \cite{sho84}. The latter can easily occur
because the scattering by magnetic impurities (or magnetic sublattices)
is in general appreciably spin dependent. The usual splitting of the
spin-up and spin-down Landau levels gives rise to a phase shift of
the oscillations by $\Phi = \pm \pi g m^*/m_0$, where $g$ is the
$g$ factor of the conduction electrons. If an antiferromagnetic
exchange interaction $B_{ex}$ is present, the phase difference
can be expressed as $\Phi = \pm\pi(g - B_{ex}/B) m^*/m_0$ \cite{sho84}.
The superposition of the spin-up and spin-down oscillations gives
a signal which is $M = a_\uparrow\sin(\psi + \Phi/2) + a_\downarrow
\sin(\psi - \Phi/2)$, with $\psi = (2\pi F/B)\pm \pi/4$ and
the spin-dependent amplitudes $a_\uparrow$ and $a_\downarrow$.
In case of largely different amplitudes, the frequency of the
oscillating signal becomes $F^\prime = F \pm B_{ex}/4$. Adopting
this scenario would, however, imply a change of $B_{ex}$ with $T$
by several 10\,T which appear unlikely. On the other hand, the
apparent decrease of the oscillation frequency in $1/B$ above
about 0.12\,T$^{-1}$ at low $T$ does suggest an influence of
internal and/or exchange fields. Calculations of spin splitting
in low-carrier-density systems are necessary to check the viability
of this scenario.

In summary, we have presented an unusual $T$ dependence of SdH
oscillations in CeBiPt which are due to the magnetic Ce
$4f$ electrons. Simple estimates of the scattering rate derived
from the Dingle temperature and of the volume enclosed by the
Fermi surface yield good agreement with the corresponding
quantities obtained from other measurements. This lends
confidence to the assignment of the observed apparent $T$
dependence being a feature associated with the magnetic quantum
oscillations in CeBiPt. However, theoretical work is needed to
unravel the origin of this new phenomenon.

We would like to thank A. Schr\"oder and J.\ Sereni
for many helpful discussions.
This work was supported by the Deutsche Forschungsgemeinschaft,
SFB 195, and the TMR Programme of the European Community under
contract No.\ ERBFMGECT950077.


\begin{references}
\bibitem{remtn}
$T_N = 2.5$\,K has been reported for polycrystalline
samples \cite{can91}.
\bibitem{remhall}
Clear oscillations were also observed in $\rho_{xy}$. The frequency of
these oscillations showed the same angular and temperature dependence
as observed in $\rho_{xx}$.
\end{references}
\end{document}